\documentclass{cjaa}                   

\usepackage{graphicx}                   
\input{epsf.sty}                        
\input{psfig.sty}                       

\newcommand{\gppr}{\stackrel{>}{\scriptstyle \sim}}
\newcommand{\lppr}{\stackrel{<}{\scriptstyle \sim}}
\newcommand{\beq}{\begin{equation}}
\newcommand{\eeq}{\end{equation}}

\begin{document}

   \title{Helical motion and the origin of QPO in blazar-type sources}

   \volnopage{Vol.0 (200x) No.0, 000--000}      
   \setcounter{page}{1}          

   \author{Frank M. Rieger
      \inst{1}\mailto{}
      }
   \offprints{Frank Rieger}                   

   \institute{Department of Mathematical Physics, University College Dublin, Belfield, Dublin 4,
              Ireland\\
             \email{frank.rieger@ucd.ie}
          }

   \date{Received~~2004; accepted~~2004}

   \abstract{Recent observations and analysis of blazar sources provide strong evidence for (i) 
    the presence of significant periodicities in their lightcurves and (ii) the occurrence of 
    helical trajectories in their radio jets. In scenarios, where the periodicity is caused 
    by differential Doppler boosting effects along a helical jet path, both of these facts
    may be naturally tied together. Here we discuss four possible driving mechanisms for the
    occurrence of helical trajectories: orbital motion in a binary system, Newtonian-driven jet 
    precession, internal jet rotation and motion along a global helical magnetic field. We point
    out that for non-ballistic helical motion the observed period may appear strongly shortened 
    due to classical travel time effects. Finally, the possible relevance of the above mentioned 
    driving mechanisms is discussed for Mkn~501, OJ~287 and AO 0235+16.
     }

   \authorrunning{F.M. Rieger}            
   \titlerunning{Origin of QPO in blazar-type sources} 

   \maketitle

\section{Introduction}\label{sect:intro}
Collimated, highly relativistic outflows (jets) have been established for many Active 
Galactic Nuclei (AGN). Among these AGN, the blazar subclass describes those radio-loud 
objects whose often highly variable continuum is dominated by non-thermal processes 
and thought to be related to relativistically beamed emission from their inner jets 
viewed almost face-on. Variability analysis of the lightcurves from these sources has 
provided strong evidence for significant periodicities on a timescale of days to years: 
While several of the well-known TeV sources, e.g. Mkn~421, Mkn~501, 3C66A and PKS 
2155-30, seem to reveal mid-term periodicity on a timescale of several tens of days in 
their optical, X-ray or TeV data (e.g., Hayashida et al.~\cite{hay98}; Lainela et 
al.~\cite{lai99}; Kranich et al.~\cite{kra01}; Osone et al.~\cite{oso01}), the optical 
lightcurves from classical sources such as BL Lac, ON 231, 3C273, OJ~287, PKS 0735+178, 
3C345 and AO 0235+16 typically suggest longterm periodicity on a timescale of several 
years (e.g., Sillanp\"a\"a et al.~\cite{sil88}; Webb et al.~\cite{web88}; Liu et 
al.~\cite{liu95}; Fan et al.~\cite{fan97,fan98}; Valtaoja et al.~\cite{val00}; Raiteri 
et al.~\cite{rai01}; Fan et al.~\cite{fan01,fan02}). Interestingly, high-resolution
kinematic studies of parsec-scale radio jets have provided strong observational
evidence for the helical motion of components, particularly in many of the above
noted objects, e.g. in BL Lac, ON 231, 3C273, 3C345, OJ~287, PKS 0735+178 (e.g., 
Zensus et al.~\cite{zen88}; Steffen et al.~\cite{ste95}; Vicente et al.~\cite{vic96}; 
Tateyama et al.~\cite{tat98}; Gomez et al.~\cite{gom99}; Kellermann et al.~\cite{kel04}). 
It appears likely that some of the observed periodic variabilities may be associated with
differential Doppler boosting due to the time-dependent change in viewing angle for 
motion along the helical path. 
 
\section{Helical motion and the possible origin of QPO}
\subsection{Differential Doppler boosting for helical motion}
Consider an emitting component with spectral index $\alpha$ moving relativistically towards 
the observer. Doppler boosting effects are then known to lead to a modulation of the observed 
flux given by 
\begin{equation}\label{modulo}
    S(\nu)=\delta^{3+\alpha}\,S'(\nu),
\end{equation} with $S'$ the spectral flux density measured in the comoving frame and $\delta$ 
the Doppler factor given by 
\beq\label{dopplerfactor}
    \delta(t)=\frac{1}{\gamma_{b}\,[1-\beta_b(t)\,\cos\,\theta(t)]}\,,
\eeq where $\theta(t)$ is the actual angle between the velocity $\vec{\beta}_b(t)=
\dot{\vec{x}}_b(t)/c\,$ of the component and the direction of the observer, and 
$\gamma_b=1/\sqrt{1-\beta_b(t)^{\,2}}$ the bulk Lorentz factor. A periodically changing
viewing angle due to regular helical motion for example, may thus naturally lead to
a periodicity in the observed lightcurves even for an intrinsically constant flux.

\subsection{Shortening of observed period for non-ballistic helical motion}
\begin{figure}[h]
   \centering
   \includegraphics[width=60mm,height=80mm]{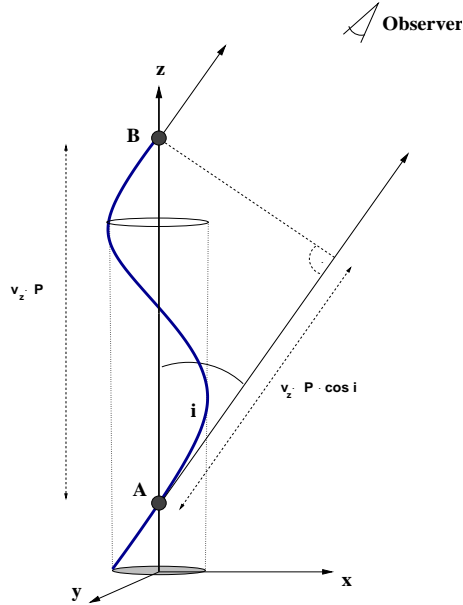}
   \caption{For non-ballistic helical motion the physical period $P$ appears shortened 
   due to classical travel-time effects, i.e. we have $P_{\rm obs} \simeq (1-[v_z/c]\,
   \cos i)\,P$.}
   \label{fig1}
   \end{figure}

It can be shown (see Rieger~\cite{rie04}) that for non-ballistic (i.e. non-radial) helical 
motion the (real) physical period $P$ appears generally shortened when measured by a distant
observer. For blazar-type sources with typical inclination angles $i \sim 1/\gamma_b$ and
bulk Lorentz factors $\gamma_b \simeq (5-15)$ the observed period is given by
\beq\label{shortening}
   P_{\rm obs} \simeq \frac{(1+z)}{\gamma_b^2}\,P\,.
\eeq This is based on the fact that observationally we measure the arrival times of pulses,
e.g. emitted at A and B (see Fig.~\ref{fig1}) where the velocity vector points closest 
towards the observer, whereas due to the relativistic motion of the emitting component
at small inclinations the light travel distance for a pulse emitted at B is much smaller
than the one for a pulse emitted at A.

\subsection{Possible periodic driving processes and associated constraints}
In general, a helical trajectory may arise due to several driving processes, including (1) 
orbital motion in a binary black hole system (BBHS), (2) Newtonian precession, (3) internal 
jet rotation or (4) global helical magnetic fields. Scenarios associated with (1) and (2) 
usually rely on the plausibility of close BBHSs in AGN. Today, BBHS are indeed expected to 
be present in the center of elliptical galaxies as a result of mergers between spiral galaxies, 
each containing its own BH (e.g., Begelman et al.~\cite{beg80}; Richstone et al.~\cite{ric98}).
A multitude of observational evidence including misalignment, precession and wiggling of jets, 
long- and short-term periodicities and X-shape radio morphologies has been successfully 
interpreted within a binary framework, whereas more direct observational support has also 
been recently established for NGC 6240 (cf., Komossa~\cite{kom03} for a review). Existing 
binary models aimed at explaining periodicity usually require very close BBHS with separation
$\lppr 10^{17}$ cm. Stability arguments (e.g., cosmological evolution, longterm periodicity) 
against loss of orbital angular momentum via gravitational radiation typically suggest 
(Keplerian) orbital periods $P_k$ of the order of several years (or larger) and hence 
observable periods $P_{\rm obs} \gppr 10$ days (cf. Eq.~\ref{shortening}) for 
{\it orbital-driven helical motion} (Rieger~\cite{rie04}). Periodicity may however 
also arise due to {\it Newtonian-driven jet precession:} Tidally induced perturbations 
in the accretion disk due to the influence of the binary companion for example, can lead to 
a rigid-body precession of the inner parts of the disk and thus translate into a precession 
of the jet (e.g., Katz~\cite{kat97}; Larwood,~\cite{lar98}; Romero et al.~\cite{rom00,rom03}). 
In general, the implied (physical) precessional periods $P_p$ are usually a factor of order 
ten times larger than the orbital period $P_k$ of the binary (Rieger~\cite{rie04}). 
Hence non-ballistic helical motion due to Newtonian jet precession is unlikely to be 
responsible for periodicity on a timescale $P_{\rm obs} \lppr 100$ days (cf. 
Eq.~\ref{shortening}), but may well be associated with periods $P_{\rm obs} \gppr 1$ yr. 
{\it Internal jet rotation} represents yet another possibility, provided components 
are dragged with the underlying rotating flow. The occurrence of such rotation (at least 
initially) appears well-supported: The strong correlation between the disk luminosity and 
bulk kinetic power in jets and the observational evidence for a jet-disk symbiosis (e.g., 
Rawlings \& Saunders~\cite{raw91}; Falcke \& Biermann~\cite{fal95}) for example, indicate 
that a significant amount of accretion energy, and hence rotational energy (cf. virial 
theorem), is channeled into the jet. Moreover, internal jet rotation is also implied in 
theoretical MHD models of jets as magnetized disk winds (e.g., Camenzind \& 
Krockenberger~\cite{cam92}; Sauty et al.~\cite{sau02}) and supported by recent observations 
of stellar jets (e.g., Coffey et al.~\cite{cof04}). Knowledge of the underlying rotation 
law may then be used to estimate possible observable periods. It can be shown (Rieger~\cite{rie04}) 
for example, that for the elaborate lighthouse model of Camenzind \& Krockenberger~(1992) 
bounds on the maximum jet radius derived from numerical simulations translate into 
characteristic periods of $P_{\rm obs} \lppr 10$ days (for massive quasars) and $P_{\rm obs} 
\sim 1$ day for typical BL Lac objects. Finally, periodicity may perhaps also be related to 
a component moving along a putative {\it global helical magnetic field}. Such a configuration 
may possibly be associated with certain types of MHD models (cf., K\"onigl \& Choudhuri
\cite{kon85}; Koide et al.~\cite{koi02}) and may observationally be suggested by rotation 
measure asymmetries across the jet in several sources (e.g., Asada et al.~\cite{asa02}; 
Gabuzda et al.~\cite{gab04}). However, to allow for significant Doppler boosting and 
modulation one then usually requires $B_z \gg B_{\Phi}$, i.e. small helix pitch angles 
$\psi$ with $\psi \lppr i$.

\section{Possible applications and relevance}
\subsection{Mkn~501}
The nearby TeV blazar Mkn~501 ($z=0.033$) attracted attention in 1997, when the source underwent
a phase of high activity becoming the brightest source in the sky at TeV energies. Among the most
interesting features during this high state are the (apparent) evidence for a periodicity on an 
observed timescale $P_{\rm obs} \simeq 23$ days, reported by several independent groups in both 
the TeV and X-ray range (e.g., Hayashida et al.~\cite{hay98}; Kranich et al.~\cite{kra01}). We 
have shown recently (cf., Rieger \& Mannheim~\cite{rie00}, hereafter RM00; cf. also De Paolis et 
al.~\cite{dep02}) that such a periodicity may be caused by orbital-driven helical motion, i.e. 
the orbital motion of the relativistic jet emerging from the less massive BH (with mass $m$) in 
a close binary system with an orbital period of the order of several years, i.e. $P_k \simeq 
(6-14)$ yrs for $\gamma_b \simeq (10-15)$ (see Eq.~\ref{shortening}). Information about the 
ratio $f$ between the maximum and minimum amplitude (here $f \simeq 8$) and the relevant spectral 
index (here $\alpha \simeq 1.2$) may be used to derive an estimate for the center-of-mass distance 
$R$ and the mass ratio of the binary (cf. Eq.~7 and 8 in RM00). One thus obtains $R \sim 10^{16}$ 
cm and  
\beq
   \left(\frac{M}{10^8\,{\rm M_{\odot}}}\right) \simeq 0.89\,\left(\frac{10}{\gamma_b}\right)\,
                                                            \left(1+\frac{m}{M}\right)^{2}\,
\eeq suggesting a possible total mass for the binary in the range $8.9 \cdot 10^7\,M_{\odot}\,
(\gamma_b/10) < (m+M) < 7.1 \cdot 10^8\,M_{\odot}\,(\gamma_b/10)$, which seems well consistent 
with estimates derived from host galaxy observations (Rieger \& Mannheim~\cite{rie03a}). 
As $P_{\rm obs}$ seems still relatively small, it may be interesting to investigate whether 
the observed periodicity may also be explainable in the lighthouse scenario proposed by 
Camenzind \& Krockenberger~(\cite{cam92}). Using $f$ and $\alpha$ as given above, it can 
be shown (cf., Rieger \& Mannheim~\cite{rie03b}) that one requires $r_0/r_{\rm L} \simeq 
100$, with  $r_L \simeq 10^{14}$ cm the light cylinder radius and $r_0$ the presumed radial 
scale where the component is injected. Consistency in the lighthouse approach generally 
suggests however that $r_0 \lppr 10\,r_{\rm L}$ (cf., Camenzind~\cite{cam96}; Fendt~\cite{fen97}). 
It seems thus unlikely, that the observed periodicity in Mkn~501 is due to such a scenario.
\subsection{OJ~287}
The BL Lac object OJ~287 ($z=0.306$) is famous for its optical long-term periodicity on a 
timescale $P_{\rm obs} \simeq 11.86$ yr (Sillanp\"a\"a et al.~\cite{sil88}; Valtaoja et 
al.~\cite{val00}). Several models have been proposed to explain the periodicity, with
accretion disk interactions in a BBHS among the most prominent ones (e.g., Lehto \& 
Valtonen~\cite{leh96}; Valtaoja et al.~\cite{val00}; Liu \& Wu~\cite{liu02}). 
Suppose for simplicity that the periodicity is caused by the secondary BH crossing the 
accretion disk around the primary BH on a non-coplanar, almost circular orbit. The physical 
orbital period is then of order $P_k = 2 \times 11.65/(1+z) \simeq 18.16$ yr. According to 
Vicente et al.~(\cite{vic96}), the VLBI jet in OJ~287 reveals evidence for non-ballistic 
helical motion of components. For a typical bulk Lorentz factor for OJ~287 of $\gamma_b 
\sim 4.4$ in the radio regime (cf., Vicente et al.~\cite{vic96}; Hughes et al.~\cite{hug98}), 
orbital-driven helical motion results in an observable period (cf. Eq.~\ref{shortening})
$P_{\rm obs, radio} \sim 1.25$ yr. Interestingly, a wavelet transform analysis of the UMRAO 
radio data for OJ~287 has revealed a period of $1.12$ yr (in the observer frame) in the 
1980s (Hughes et al.~\cite{hug98}). It seems thus very tempting to relate this periodicity 
to the orbital-driven, non-ballistic helical motion of a component dominating the emission. 
Further research appears important to assess the plausibility of such a scenario.
\subsection{AO 0235+16}
The low energy peaked BL Lac object AO 0235+16 ($z=0.94$) is well known for its extreme 
variability at almost all wavelengths. Recently, the analysis of its long-term 
variability has provided evidence for a possible $\sim 5.7$ yr periodicity in its radio 
light curves and $\sim 2.95$ yr periodicity in its optical light curves (e.g., Raiteri et 
al.~\cite{rai01}; Fan et al.~\cite{fan02}). Romero, Fan \& Nuza~(\cite{rom03}) have shown 
that the optical periodicity (corresponding to $P_k \simeq 2 \times 2.95/[1+z] \simeq 3$ yr) 
might be related to the secondary crossing the accretion disk around the primary on a 
non-coplanar circular orbit, while the radio periodicity might be associated with Newtonian 
jet precession. As argued above, the ratio of the precessional to orbital period is usually 
expected to be of the order of ten or larger, i.e. $P_p \gppr 30 $ yr. Eq.~(\ref{shortening}) 
then suggests that we require bulk Lorentz factors $\gamma_b \gppr 3$. Estimates derived from 
host galaxy observations of BL Lacs usually indicates central masses in the range $6 \cdot 
10^7\,M_{\odot} \lppr (M+m) \lppr 10^9\,M_{\odot}$ (e.g., Wu et al.~\cite{wu02}; Falomo et 
al.~\cite{fal03}), thus suggesting a binary separation for AO 0235+16 likely to be in the 
range $10^{16}\,{\rm cm}\,\lppr d \lppr 3 \cdot 10^{16}\,{\rm cm}$. 

\section{Conclusions}
In this contribution we have shown that periodicity in blazar-type sources may arise
simply due to differential Doppler boosting effects along a helical path inclined at
small viewing angle. Accordingly, the observation of non-ballistic helical motions in 
radio jets and the detection of periodicities on timescales of several years or less 
(in the frequency-domain where the observed emission is dominated by the jet) may 
mutually support each other. As different driving mechanisms are usually associated
with different driving periods, the observed timescale of periodicity may be used to 
single out the most likely underlying driving mechanism, thus allowing to draw 
valuable conclusions on the nature of the central engine.


\begin{acknowledgements}
Discussions with J.H.~Fan, G.~Romero, M.~Tagger and K.~Mannheim are gratefully 
acknowledged. FMR acknowledges financial support through a Marie-Curie Individual 
Fellowship (MCIF-2002-00842).
\end{acknowledgements}

\label{lastpage}

\end{document}